\documentstyle[12pt,aaspp4,psfig]{article}

\newcommand \hii  {H\,{\sc ii}}
\newcommand \heii  {He\,{\sc ii}}

\newcommand \ha   {H$\alpha$}
\newcommand \kms  {km~s$^{-1}$}
\newcommand \vhel {V$_{\rm hel}$}
\newcommand \nii  {[N\,{\sc ii}]}
\newcommand \oiii {[O\,{\sc iii}]}

\begin{document}

\title{Physical Structure of Small Wolf-Rayet Ring Nebulae}

\author{You-Hua Chu\altaffilmark{1}, Kerstin Weis\altaffilmark{1, 2}}
\affil{Astronomy Department, University of Illinois, 1002 W. Green Street,
Urbana, IL 61801 \\
Electronic mail: chu@astro.uiuc.edu, kweis@platon.ita.uni-heidelberg.de}

\author{Donald R. Garnett\altaffilmark{1, 3}}
\affil{Astronomy Department, University of Minnesota,
116 Church Street, S. E., Minneapolis, MN 55455 \\
Electronic mail: dgarnett@as.arizona.edu}

\altaffiltext{1}{Visiting astronomer, Cerro Tololo Inter-American 
Observatory, National Optical Astronomy Observatories, operated by 
the Association of Universities for Research in Astronomy, Inc., 
under a cooperative agreement with the National Science Foundation.}

\altaffiltext{2}{Current address: Institut f\"ur Theoretische Astrophysik, 
Tiergartenstr. 15, D-69121 Heidelberg, Germany.}

\altaffiltext{3}{Current address: Steward Observatory, University of
Arizona, Tucson, AZ 85721.}

\begin{abstract}

We have selected the seven most well-defined WR ring nebulae 
in the LMC (Br 2, Br 10, Br 13, Br 40a, Br 48, Br 52, and Br 100)
to study their physical nature and evolutionary stages.
New CCD imaging and echelle observations have been obtained
for five of these nebulae; previous photographic imaging
and echelle observations are available for the remaining two 
nebulae.  Using the nebular dynamics and abundances, we find 
that the Br 13 nebula is a circumstellar bubble, and that the 
Br 2 nebula may represent a circumstellar bubble merging with a
fossil main-sequence interstellar bubble.  The nebulae around
Br 10, Br 52, and Br 100 all show influence of the ambient
interstellar medium.  Their regular expansion patterns
suggest that they still contain significant amounts of 
circumstellar material.  Their nebular abundances would be 
extremely interesting, as their central stars are WC5 and 
WN3-4 stars whose nebular abundances have not been derived
previously.  Intriguing and tantalizing implications are
obtained from comparisons of the LMC WR ring nebulae with 
ring nebulae around Galactic WR stars, Galactic LBVs, LMC
LBVs, and LMC BSGs; however, these implications may be
limited by small-number statistics.
A SNR candidate close to Br 2 is diagnosed by its large 
expansion velocity and nonthermal radio emission.
There is no indication that Br 2's ring nebula interacts 
dynamically with this SNR candidate.

\end{abstract}

\keywords{stars: Wolf-Rayet - stars: mass-loss - ISM: bubbles -
ISM: kinematics and dynamics -  Magellanic Clouds}

\clearpage

\section{Introduction}

Wolf-Rayet (WR) stars are characterized by broad emission lines 
that are indicative of fast stellar winds and high mass loss rates.
The fast stellar wind sweeps up the ambient medium into a dense 
expanding shell, called a wind-blown bubble.  Wind-blown bubbles 
in a homogeneous medium have been modeled by, for example, 
\markcite{CMW75}Castor, McCray, \& Weaver (1975), 
\markcite{SSW75}Steigman, Strittmatter, \& Williams (1975), and 
\markcite{We77}Weaver et al.\ (1977).  However, these models cannot 
be readily applied to bubbles blown by WR stars because WR stars 
are evolved massive stars and their progenitors' mass loss has 
drastically modified their gaseous surroundings.  The physical 
conditions of a WR star's ambient medium are highly dependent on 
the star's evolutionary history and mass loss history.

WR stars are divided into WN (nitrogen) and WC (carbon-oxygen) 
sequences; the WN sequence is further divided into excitation 
classes WN2--9 and the WC sequence into WC4--9 
(\markcite{Hu81}van der Hucht et al.\ 1981).  These diverse 
types of WR stars have different initial masses and 
evolutionary paths (\markcite{La94}Langer et al.\ 1994).  It 
is conceivable that ring nebulae around WR stars must have a 
variety of origins and physical conditions.

The formation of WR bubbles has been calculated both 
analytically (\markcite{GM95a}Garc\'{\i}a-Segura \& Mac Low
1995a) and numerically (\markcite{GM95b}Garc\'{\i}a-Segura \&
Mac Low 1995b).  Hydrodynamic models of bubbles have been
produced for WR stars descendent from luminous 
blue variables (LBVs; \markcite{GML96}Garc\'{\i}a-Segura, 
Mac Low, \& Langer 1996, hereafter GML96) and from red 
supergiants (RSGs; \markcite{GML96}Garc\'{\i}a-Segura, Langer,
\& Mac Low 1996, hereafter GLM96), respectively.
These models conclude that the abundances, morphology, and 
kinematics of a WR bubble can be used to diagnose the 
evolutionary status of the bubble and to determine whether the 
central WR star is a descendant of a LBV (e.g., RCW 58) or 
a RSG (e.g., NGC\,6888).

It is of interest to compare these models to a large number 
of WR bubbles to determine the nature of the bubbles and the 
applicability of the models.  Recent CCD surveys have found 
many new WR ring nebulae in our Galaxy (\markcite{MC93}Miller 
\& Chu 1993; \markcite{Ma94a}\markcite{Ma94b}Marston et al.\ 
1994a, 1994b) and in the Large Magellanic Cloud (LMC; 
\markcite{Do94}Dopita et al.\ 1994).  These ring nebulae 
provide excellent samples of WR bubbles for comparisons 
with models.  

We have begun to study the LMC sample because the extinction 
is small, the distance is well-known, and the LMC WR bubbles 
can be easily studied with long-slit spectrographs.  Table 1 
lists the most well-defined small WR ring nebulae in the LMC.  
The WR stars Br 2, Br 10, Br 13, Br 48, Br 52, and Br 100 were 
from \markcite{Br81}Breysacher (1981), and the star Br 40a 
was from \markcite{CG83}Conti \& Garmany (1983).  We have 
obtained high-dispersion echelle/CCD observations for five of 
these small WR rings to determine their dynamic structures and 
to diagnose their nature.  In this paper we report the 
observations (\S 2), interpret the data (\S 3), and compare 
the LMC WR ring nebulae to the hydrodynamic models of 
GML96 and GLM96, and compare them to the ring nebulae around
Galactic WR stars, LBVs, and blue supergiants (\S 4).  
A summary is given at the end (\S 5).

\section{Observations}

\subsection{Imaging}
 
We have obtained emission-line images of five small WR ring 
nebulae in the LMC using CCD cameras on the 0.9 m telescope at 
Cerro Tololo Inter-American Observatory (CTIO).  In December 
1991, the nebulae around Br 2 and Br 13 
(\markcite{Br81}Breysacher 1981) were imaged in the \ha\ and 
\oiii\,$\lambda$5007 lines with a Tektronix 1024$\times$1024 
CCD (Tek1024\#1).  The pixel size was 0\farcs4 pixel$^{-1}$ 
and the field of view was 6\farcm8$\times$6\farcm8.  In 
January 1996, the nebulae around Br 40a, Br 48, and Br 52 
were imaged in the \oiii\,$\lambda$5007 line with a Tektronix 
2048$\times$2048 CCD (Tek2K3).  The pixel size was 0\farcs4 
pixel$^{-1}$ and the field of view was 13\farcm5$\times$13\farcm5.  

The exposure time for each image was typically 900 s.  The ring 
nebula around Br 13 is so faint that two 900 s exposures were 
averaged together to produce the final image in each filter. 
Images of the five LMC WR ring nebulae are presented in Figure 1.
Note that the \ha\ images include contributions of the 
\nii\ $\lambda\lambda$6548, 6583 lines, as the \ha\ filter was 
centered at 6575 \AA\ with a FWHM of 14 \AA.  Using the 
line intensities measured by Garnett \& Chu (1994), we find that 
the \nii\ contribution to the \ha\ images is $<$1.5\% for the 
Br 2 nebula and $<$7\% for the Br 13 nebula.

\subsection{Echelle Spectroscopy}
 
High-dispersion spectroscopic observations of five LMC WR 
rings were obtained with the echelle spectrograph on the 4 m 
telescope at CTIO in January 1996.  The spectrograph was used 
in a long-slit mode by inserting a post-slit \ha\ filter 
(6563/75 \AA) and replacing the cross-disperser with a flat 
mirror.  A 79 lines mm$^{-1}$ echelle grating and the long 
focus red camera were used.  The detector was a Tektronix 
2048$\times$2048 CCD (Tek2K4).  The pixel size was 0.08 \AA\ 
pixel$^{-1}$ along the dispersion and 0\farcs26 pixel$^{-1}$ 
in the spatial axis.  The slit length was effectively limited
by vignetting to $\sim4'$.  Both \ha\ $\lambda$6563 and 
\nii\ $\lambda\lambda$6548, 6583 lines were covered 
in this setup.  The slit-width was 250 $\mu$m, or 1\farcs64,
leading to an instrumental FWHM of $\sim$14 \kms\ at the 
\ha\ line.  Thorium-Argon lamp exposures were obtained
for wavelength calibration and geometric distortion correction.

The journal of echelle observations is given in Table 2, and 
the echellograms of the \ha\ and the \nii\ $\lambda$6583 lines 
are shown in Figure 2.  Each panel covers 38.48 \AA\ along the 
dispersion (the horizontal axis), and 2$'$ for Br 2 and Br 13
and 4$'$ for Br 40a, Br 48, and Br 52 along the slit (the 
vertical axis).  The narrow, unresolved lines
are telluric \ha\ and OH lines (Osterbrock et al.\ 1996),
which provide convenient references for fine-tuning the 
wavelength calibration.  The OH 6-1 P2(3.5)6568.779 line is
unfortunately blended with the LMC \ha\ line.  This OH line, 
corresponding to the \ha\ line at a heliocentric velocity of 
V$_{\rm hel}\sim275$ \kms, can be seen on the left 
side of the \ha\ line of the Br 52 nebula (Figure 2).

\section{Physical Structure of the Five WR Rings}

\subsection{Br 2}

The ring nebula around Br 2 was at first identified to be 
the 3$'$ shell nebula to the south of Br 2 (Rosado 1986).  
It was later discovered in higher resolution images that 
a 28$''\times18''$ arc surrounds Br 2 (Dopita et al.\ 1994).
This small arc has a morphology similar to that of 
NGC\,6888, but has a much higher excitation and emits
\heii\ $\lambda$4686 line (\markcite{Pa91}Pakull 1991).  
The contrast between this arc and the background \hii\ 
region is higher in \ha\ images than in \oiii\ images 
(see Figure 1).

We assume that the small arc is the bona fide ring nebula 
of Br 2.  The echelle observations along position angles 
PA = 120$^\circ$ and 135$^\circ$ (near the minor axis) show
a broader \ha\ line within the arc; the FWHM of the \ha\ 
line is 42$\pm1$ \kms\ near Br 2, and 32$\pm1$ \kms\ outside
the arc.  The centroid of the \ha\ line near Br 2 is 
blue-shifted by 3 \kms\ with respect to the background \hii\ 
region velocity at \vhel\ = 253$\pm$2 \kms.  This small 
velocity shift prohibits an unambiguous decomposition of 
the broadened line into multiple components.  Fortunately, 
the \heii\ $\lambda$6560.18\footnote{We have calibrated the 
wavelength of the \heii\ line using echelle observations 
of the \hii\ region N44C, of which the \ha\ and \heii\ 
lines are narrow.  We find the \heii\ line to be at 
2.60$\pm$0.05 \AA\ shorter wavelength than the \ha\ line.  
For an \ha\ wavelength of 6562.78 \AA, the \heii\ line 
will be at 6560.18 \AA.}
line is detected within and beyond the arc.  The \heii\ 
line is less confused by the background \hii\ region 
emission, hence provides better kinematic diagnostics
for Br 2's ring. 

The \heii\ line is curved.  In the background \hii\ 
region exterior to the arc, the centroid velocity of the 
\heii\ line, 250$\pm$3 \kms, is similar to that of the 
\ha\ line.  Near Br 2 the \heii\ line is blue-shifted 
by 16$\pm$2 \kms\ with respect to the \heii\ velocity 
exterior to the arc.  A similar \heii\ velocity structure 
is seen in the echellogram taken along PA = 45$^\circ$ near
the major axis the arc.  This velocity structure indicates 
that the stellar wind of Br 2 has accelerated the ambient 
medium by 16 \kms.  The accelerated ambient medium is 
probably interstellar matter, as it has normal LMC 
interstellar abundances (Garnett \& Chu 1994).

It is worth noting that a supernova remnant (SNR) candidate
is detected to the northwest of Br 2's ring along PA =
120$^\circ$ and 135$^\circ$.  This SNR candidate appears 
as a high-velocity feature, with velocity offsets up to 150 
\kms, extending from the Br 2's ring to the northwest for at 
least 80$''$.  Our long-slit low-dispersion spectra of this 
region also show enhanced [S\,{\sc ii}] and [O\,{\sc i}] 
lines characteristic of SNRs (Garnett \& Chu 1994).

\subsection{Br 13}

The ring nebula of Br 13 is elongated along the position
angle 33$^\circ$.  Its semi-major axis is 20$''$ long
and its semi-minor axis 16$''$, corresponding to 5 pc 
and 4 pc, respectively.  Only the northwestern half of 
the ring is detected in our \ha\ and \oiii\ images. 

Our echelle observation along the E--W direction shows a
stationary component from the background \hii\ region and
a velocity-position ellipse from the ring nebula.  The 
velocity-position ellipse, indicating an expanding shell,
is detected in both the \ha\ and \nii $\lambda$6583 lines,
although the latter has a much lower S/N ratio. As in the 
direct images, the surface brightness of the expanding shell 
is not uniform.  The receding side of the shell is detected 
for 19$''$ on the west side of Br 13 and $\sim$20$''$ on 
the east, but the approaching side is detected only on the 
west side.  The largest velocity split, 160 \kms, is 
detected near the shell center.  

To determine the expansion velocity, we need to know the 
systemic velocity of the nebula.  Br 13 and its ring
nebula are projected against the outskirts of the \hii\ 
region DEM\,L\,56 (\markcite{DEM}Davies, Elliott, \& 
Meaburn 1976) which has a heliocentric velocity of 304 \kms.
However, this velocity does not necessarily represent the 
velocity of Br 13's ring nebula because the ring nebula 
consists of ejected stellar material (Garnett \& Chu 1994) 
and Br 13 may not have the same origin and systemic velocity 
as DEM\,L\,56.  The stellar \heii\ $\lambda$6560 line profile 
of Br 13, having a  FWHM of $\sim$1080 \kms\ and containing 
an unknown amount of stellar \ha\ emission, cannot be used 
to determine the radial velocity of Br 13.  We have to resort
to the nebular velocities to assess the systemic velocity
of Br 13 and its ring nebula.

The echelle data show that the western tip of the ring 
nebula's velocity-position ellipse converges to the \hii\ 
region velocity, but the eastern tip is not detected.  The 
faint velocity-position ellipse in the [N\,{\sc ii}] line 
appears to be slightly tilted, with the western end toward 
lower velocity and the eastern end higher velocity.  Since 
the [N\,{\sc ii}] line is detected at a very low S/N, it 
is difficult to determine the exact amount of line tilt.
If the expansion is symmetric with respect to the central
star, the radial velocities of the approaching and receding 
sides of the shell at the center, 237 and 397 \kms, would
imply a systemic velocity of 317 \kms\ and an expansion 
velocity of 80 \kms\ along the line of sight toward Br 13.
Br 13's ring nebula would be moving at +13 \kms\ relative 
to the background \hii\ region.  This relative motion may 
be responsible for the higher surface brightness of the
leading side of the shell.  Nevertheless, we cannot 
confidently rule out the possibility that Br 13 and its 
ring nebula do have the same systemic velocity as the 
\hii\ region DEM\,L\,56, and the expansion of Br 13's 
circumstellar bubble is asymmetric, with the expansion 
velocity being 67 \kms\ on the approaching side and 93 
\kms\ on the receding side.

\subsection{Br 40a}

Br 40a is located in the northwestern quadrant of the \hii\ 
complex N206 (\markcite{He56}Henize 1956), or DEM\,L\,221.  
\ha\ images show a parabolic-shaped arc around Br 40a, which 
has been suggested to be a ring nebula (Dopita et al.\ 1994).
However, our \oiii\ image shows that the \ha\ arc is composed
of two filaments with different excitations.  The eastern part
of the \ha\ arc has such a low excitation that it is not detected
in our \oiii\ image.  The shape of the \ha\ arc is probably 
fortuitous.

Our echelle observation has a N--S oriented slit passing 
through Br 40a.  Within the nebula around Br 40a, no line 
split is detected, but the centroid of the \ha\ line is 
red-shifted by several \kms\ with respect to that of the
background \hii\ region.  The \ha\ velocity varies from 
232 \kms\ southward of Br 40a's ring to 242--245 \kms\
within the boundary of Br 40a's ring, and to 237 \kms\
at large distances to the north.  A He\,{\sc ii}-emission
region around Br 40a has been reported by Niemela (1998);
however, our echelle observation did not detect the 
He\,{\sc ii}$\lambda$6560 line.

The velocity pattern of Br 40a's ring nebula does not 
suggest an expanding shell, but does suggest interactions 
between Br 40a and its surrounding medium.  The anomalous 
velocity in the small nebula round Br 40a is probably 
caused by the fast stellar wind accelerating the dense
interstellar medium on the far side of the star.

\subsection{Br 48}

The ring nebula around Br 48 is inside the \hii\ region
DEM\,L\,231.  The ring has a ``double rim" morphology that
resembles a tilted short cylinder.  This distinct morphology
is remarkably reminiscent of the Ring Nebula, although their 
nature and dynamic structures are completely different.  
Br 48's ring has a dimension of 95$''\times70''$, or 
24$\times$17 pc, and the surrounding \hii\ region is 
about 50 pc in diameter.

The velocity field of DEM\,L\,231 has been previously 
studied with the same echelle spectrograph on the same 
telescope but using photographic plates (\markcite{Ch83}Chu 
1983).  Based on those photographic data, it was deduced 
that Br 48 interacts with a tilted slab of interstellar gas 
and that no three-dimension expanding shell is present.  
This basic picture is still supported by our new 
observations using a more sensitive detector.

Our new echelle observation samples a E--W cut through the 
central star.  An apparent velocity gradient is detected
in the \hii\ region, with the heliocentric velocity varying 
from 311 \kms\ on the east to 297 \kms\ on the west. 
Within the central cavity, the \ha\ line is centered at 
295--300 \kms.  Relative to this central velocity, 
the inner rim of Br 48's ring is blue-shifted at
\vhel\ $\sim$ 294 \kms\ on the east side, and red-shifted at
\vhel\ $\sim$ 306 \kms\ on the west side.  This indicates 
that Br 48 indeed interacts with a slab of gas and that 
this slab is tilted with the east side toward us.

No high-velocity components are detected near the central
cavity.  The apparent velocity FWHM at the central cavity 
reaches 45 \kms, as opposed to 30$\pm$1 \kms\ in the 
brighter and more quiescent parts of the ring or the \hii\ 
region.  Quadratically subtracting a thermal FWHM of 21 
\kms\ (for 10$^4$ K) and an instrumental FWHM of 14 \kms, 
we derive a turbulent FWHM of 37 \kms\ for the central 
cavity and 16 \kms\ in the bright (quiescent) region.
If the broadening of velocity profiles at the central 
cavity is caused by an expansion, the expansion velocity 
cannot be larger than 18$\pm$1 \kms.


\subsection{Br 52}

The ring nebula of Br 52 was once thought to be the 
40$''$ triangular-shaped nebula to the east of the star 
(\markcite{CL80}Chu \& Lasker 1980).  Recent CCD images 
have detected the fainter part of the ring nebula to the 
southwest of Br 52 (Dopita et al.\ 1994).  The \oiii\ 
image in Figure 1 shows a complete shell structure, 
although the southwestern half is fainter.  

Our new echelle observation has a E--W oriented slit 
centered on Br 52.  The \ha\ line is dominated by one
bright component.  The heliocentric velocity of this 
main component is 320 \kms\ outside the ring nebula.
Within the ring nebula, the velocity of the main \ha\ 
component is marginally blue-shifted, by 2 \kms, on the 
west side of Br 52 and red-shifted by up to 5 \kms\ on 
the east side of Br 52.  An additional faint, red-shifted
\ha\ component is present within the ring nebula on the 
west side of Br 52.  The heliocentric velocity of this 
component reaches 370 \kms, which is offset from the main 
component by +50 \kms.

The velocity structure of Br 52's ring nebula indicates 
a ``blister" structure.  This ring nebula is most likely 
an interstellar bubble blown by Br 52 in a medium with
a steep density gradient.  The bubble/blister expands 
into the lower-density medium with an V$_{\rm exp}$ of 
$\sim$50 \kms.

\section{Discussion}

The formation of a WR ring nebula can be qualitatively
described by the following scenario (GML96; GLM96).  A 
massive star evolves off the main sequence, passes through 
a RSG phase or a LBV phase, then lands on a WR phase.  During 
the main sequence stage, the fast stellar wind sweeps up the
ambient interstellar medium to form an interstellar bubble.
The copious mass loss during the RSG or LBV phase would form a 
circumstellar envelope within the central cavity of the main 
sequence bubble.  As the central star evolves into a WR star,
the fast WR wind compresses the circumstellar envelope of 
previous RSG or LBV wind to a dense shell, forming a 
circumstellar bubble.  As the circumstellar bubble expands
past the outer edge of the circumstellar envelope, 
instabilities set in and the dense shell fragments.
The circumstellar bubble may collide and merge with the main
sequence interstellar bubble or evaporate in the hot shocked 
WR wind in the interior of the main sequence bubble.

An observed WR ring nebula could be in any of these 
aforementioned stages.  It is conceivable that interstellar 
bubbles have large dynamic ages and normal (interstellar)
abundances, while the circumstellar bubbles have small 
dynamic ages and anomalous abundances.  As a circumstellar 
bubble merges with the surrounding interstellar bubble, 
the observed abundances will asymptotically approach the 
normal (interstellar) abundances.  The most telltale physical 
properties that may distinguish these stages are thus nebular 
dynamics and abundances.  However, it must be born in mind
that large variations exist in the history of stellar mass
loss and the distribution of ambient interstellar medium;
therefore, even detailed information on dynamics and 
abundances may not lead to a unique interpretation.

We are extending the study of WR ring nebulae to the LMC
sample and have selected the seven most well-defined WR 
rings in the LMC from the survey by Dopita et al.\ (1994).
These seven ring nebulae, listed in Table 1, not only have 
distinct ring or arc morphology but also have sizes in the 
range of 5 to 40 pc, so that these ring nebulae may be the 
true counterparts of the archetypical WR rings, such as 
NGC\,2359, NGC\,6888, and S\,308 in the Galaxy (Johnson \& 
Hogg 1965).  

We will first discuss the nature of these seven LMC WR ring 
nebulae and compare them to the Galactic WR ring nebulae.
We will further compare the WR ring nebulae to ring nebulae
around LBVs and blue supergiants (BSGs).  As these different
spectral types represent different evolutionary stages of
massive stars, the comparison of physical properties among
their ring nebulae may help us understand the evolutionary 
aspects of these stars.  The size, expansion velocity,
``dynamic timescale", and N/O abundance ratios of these 
nebulae are tabulated in Table 3.  The ``dynamic timescale"
is defined as the expansion velocity divided by the radius.
This dynamic timescale scales with the dynamic age of the 
nebula, but is not equal to the dynamic age which depends
on whether the expansion has been accelerated or decelerated
since the initial formation.

\subsection{LMC and Galactic WR Ring Nebulae}

The kinematic structure of each of the seven selected LMC WR 
ring nebulae (Chu 1983; this paper) suggests interactions 
between the fast stellar wind and the ambient medium.  In some 
nebulae, especially those around Br 40a and Br 48, the 
kinematic structure is so irregular that it cannot be 
approximated as an expanding shell.  These irregular motions 
are clearly caused by the large density variations in the 
ambient interstellar medium; consequently, the kinematic 
structure of these nebulae cannot be used to derive 
unambiguous information on the stellar mass loss history.
For the rest of this discussion we will concentrate on only 
the other five nebulae that have well-behaved expansion 
properties.

For comparison, we have selected six Galactic WR ring nebulae, 
listed in Table 3, based on their well-observed dynamics and 
abundances.  Relative to a Galactic interstellar N/O ratio of 
$\sim$0.07$\pm$0.01 (Shaver et al.\ 1983), S308, RCW\,58, 
M\,1-67, and NGC\,6888 are obviously enriched and must contain
stellar ejecta.  NGC\,2359 and NGC\,3199, on the other hand, 
show little abundance anomaly, indicating that these nebulae
are dominated by interstellar material.  These two nebulae 
also have the most irregular expansion patterns and the 
smallest expansion velocities.  These dynamic properties 
and N/O abundance ratios suggest that the circumstellar
bubbles of NGC\,2359 and NGC\,3199 have merged with 
the fossil main sequence interstellar bubbles.  Using the 
detailed fragmentation morphology, GLM96 and GML96 conclude 
that NGC\,6888's WR star has evolved through a RSG phase, 
and RCW\,58's WR star has evolved through a LBV phase.

The LMC WR ring nebulae do not have as many observations of 
abundances available as the Galactic nebulae.  Only two LMC
nebulae, around Br 2 and Br 13, have been observed (Garnett 
\& Chu 1994).  Compared to the LMC interstellar value of N/O 
$\sim$0.04 (Garnett 1998), the Br 13 nebula has clearly 
anomalous abundances while the Br 2 nebula is marginally 
anomalous.  The Br 13 nebula must be a circumstellar bubble; 
its regular expansion pattern and small dynamic timescale both 
support this explanation.  The spectral type of Br 13, WN8, is 
the same as that of WR40, the central star of RCW\,58; the N/O 
ratio of the Br 13 nebula is similar to that of RCW\,58.  If 
the progenitor of WR40 was a LBV, it is then possible that the
progenitor of Br 13 was also a LBV.  A high-resolution image 
of the Br 13 nebula would be useful in determining whether the
fragmentation of the nebula is consistent with those expected
for a LBV progenitor (GML96).  The N/O ratio of the Br 2 nebula 
indicates that it might be a circumstellar bubble merging with 
a main sequence interstellar bubble.  This explanation is 
supported by the large dynamic timescale and small expansion 
velocity of the Br 2 nebula.

The other three LMC WR ring nebulae, around Br 10, Br 52, and 
Br 100, do not have abundance observations.  Their large sizes
and surface brightness variations indicate that they must be
interacting with the ambient interstellar medium.  However,
these three nebulae have relatively large expansion velocities
and quite regular expansion pattern, especially the Br 10 nebula.
If the LMC WR nebulae behave similarly to the Galactic WR nebulae,
we may expect that these three nebulae are not yet dominated by 
interstellar material.  The spectral types of these three
central stars are WC5 and WN3-4.  No abundances have been 
derived for circumstellar nebulae of WC5 or WN3-4 stars
in either the Galaxy or the Magellanic Clouds.  New abundance
observations of these three WR nebulae would be most interesting,
as they could place constraints on the evolution of progenitors
for these spectral types.

\subsection{Comparison with Ring Nebulae around Other Massive Stars}

We have included in Table 3 a number of Galactic and LMC LBV 
nebulae whose nebular dynamics have been well observed.  The 
Galactic LBV nebulae are all smaller and younger than the 
Galactic WR ring nebulae.  Among the Galactic LBV nebulae, the 
smallest one has the smallest dynamic timescale and the lowest 
expansion velocity.  If these four Galactic LBVs evolve similarly,
their nebulae must have gone through a rapid acceleration during 
the LBV phase. 

The LMC LBV nebulae appear to be generally larger than the 
Galactic LBV nebulae. This could be completely caused by 
a combination of small-number statistics and difficulty in
resolving and detecting small LBV nebulae in the LMC where
1 pc subtends only 4$''$.  It is noteworthy, however, that 
the LMC LBV nebulae have smaller expansion velocities than 
the Galactic counterparts.  This relation, if it holds for
a larger number of LMC LBV nebulae, may indicate a difference 
in the mass loss properties between the LMC LBVs and the 
Galactic LBVs.  

It is interesting to compare the Br 13 nebula to the LMC LBV
nebulae.  The dynamic timescale of the Br 13 nebula is 
comparable to those of the LMC LBV nebulae of R127 and S119,
although the Br 13 nebula is much larger and expands much
faster.  This comparison suggests that the progenitor of 
Br 13 could not have been a LBV similar to R127 or S119.

Finally, we examine two ring nebulae around LMC blue 
supergiants (BSGs).  The ring nebula around the O9f star 
Sk$-$69\,279 is recently discovered by Weis et al.\ (1997b).  
Its high \nii$\lambda$6583/\ha\ ratio, $\sim$0.7, is 
consistent with that expected in a N-enriched ejecta
nebula.  Its expansion velocity, only 14 \kms, is lower 
than those of all WR and LBV nebulae listed in Table 3.
Its size is larger than that of every known LBV nebula,
but comparable to those of small WR ring nebulae. 
The exact abundances of Sk$-$69\,279's nebula are unknown,
hence it is uncertain whether the nebula has swept up a 
significant amount of interstellar material and its
expansion has subsequently been slowed down.  Future 
abundance observations are needed to determine the 
evolutionary status of Sk$-$69\,279 with respect to RSG,
LBV, and WR phases.  The ring nebula around the B3I star
Sk$-$69\,202, better known as the progenitor of SN 1987A,
is small compared to ring nebulae around other massive stars.
Its expansion velocity is the smallest.  Clearly, the size 
and dynamics of Sk$-$69\,202's ring nebula suggest that 
Sk$-$69\,202 could not have gone through a WR phase.  This 
is consistent with the relatively low mass (20 M$_\odot$)
inferred for the supernova SN 1987A's progenitor.

\subsection{A SNR Candidate Near Br 2}

The echelle observations of Br 2 reveal a high-velocity 
feature to the northwest of Br 2.  The echellogram of
the slit position centered on Br 2 along the position 
angle 120$^\circ$ shows high-velocity material projected 
from the vicinity of Br 2 to almost 90$''$ northwest of 
Br 2.  The slit position centered at 10$''$N, 35$''$W of
Br 2 along the N-S direction shows high-velocity material
over 110$''$, or 28 pc, along the slit (Figure 2).  The 
velocity structure and size are very similar to those of 
known SNRs in the Magellanic Clouds, particularly the SNR N19
(0045$-$73.4) in the Small Magellanic Cloud (\markcite{CK88}Chu 
\& Kennicutt 1988).  The radio continuum emission of this 
region is brighter than those of \hii\ regions of comparable 
or even higher H$\alpha$ surface brightnesses 
(\markcite{Ha91}Haynes et al.\ 1991), indicating a 
nonthermal radio emission.  Thus, this high-velocity 
feature most likely originates from a SNR at an age of a 
few $\times 10^4$ yr.

Br 2's ring nebula is unusual in two respects.  First,
He\,{\sc ii} emission is detected in the ring nebula.
Second, Br 2's ring overlaps the projected position of 
a SNR candidate.  However, the He\,{\sc ii} emission
cannot be caused by a dynamic interaction between the 
ring nebula and the SNR, as the He\,{\sc ii}$\lambda$6560 
line does not show violent velocities and the H$\alpha$
line of the ring nebula does not show continuous
high-velocity wings.  The SNR is probably just projected
by chance to the vicinity of Br 2 and its ring nebula.

\section{Summary}

We have selected the seven most well-defined WR ring nebulae
in the LMC to study their physical nature and evolutionary
stages.  New images and echelle observations have been
obtained for five of these nebulae; previous observations
(Chu 1983) are available for the two remaining nebulae.  
Only five of these nebulae (Br 2, Br 10, Br 13, Br 52, and 
Br 100) have well-behaved expansion pattern to warrant 
further discussion.  Of these five nebulae, Br 2's and 
Br 13's ring nebulae have abundance information available
(Garnett \& Chu 1994).

Based on nebular dynamics and abundances, we suggest that the
Br 13 nebula is a circumstellar bubble, and that the Br 2 
nebula may represent a circumstellar bubble merging with a
fossil main-sequence interstellar bubble.  The nebulae around
Br 10, Br 52, and Br 100 all show influence of the ambient
interstellar medium.  Their regular expansion patterns
suggest that they still contain significant amounts of 
circumstellar material.  The abundances of these nebulae
would be extremely interesting, as their central stars 
are WC5 and WN3-4 stars whose nebular abundances have
never been derived before.

The LMC WR ring nebulae do not differ significantly from
their Galactic counterparts.  Comparisons between WR ring 
nebulae and ring nebulae around other massive stars, such 
as LBVs and BSGs, yield intriguing and tantalizing 
implications on stellar evolution and mass loss history.
However, the credibility of these implications is limited
by a small number statistics.  Future observations of a 
larger number of nebulae are needed to confirm these results.

\acknowledgments YHC acknowledges the support of NASA LTSA 
grant NAG 5-3246.  DRG acknowledges support from NASA LTSA 
grant NAG 5-6416.

\clearpage
\vskip 50pt

\begin{table}[h]
\caption[junk]{Small Wolf-Rayet Ring Nebulae in the 
Large Magellanic Cloud}
\renewcommand{\footnoterule}{}
\begin{minipage}{5.75in}
\begin{tabular}{llc}
\hline \hline
 WR$^a$  & Nebula$^b$     &   Size        \\
Star     &   Name         & (arcsec)      \\
\hline
 Br 2    & in DEM\,L\,6   & 28$\times$18   \\
 Br 10   & in DEM\,L\,39  & 190$\times$100 \\
 Br 13   & in DEM\,L\,56  & 41$\times$22   \\
 Br 40a$^c$  & in DEM\,L\,221 & 75$\times$40  \\
 Br 48   &  DEM\,L\,231   & 95$\times$70   \\
 Br 52   &  DEM\,L\,240   & 100$\times$70  \\
 Br 100  &  DEM\,L\,315   & 125$\times$90  \\
\hline
\end{tabular}

Notes:  \\
$^a$ \markcite{Br81}Breysacher 1981. \\
$^b$ \markcite{DEM76}Davies, Elliott, \& Meaburn 1976. \\
$^c$ \markcite{CG83}Conti \& Garmany 1983.
\end{minipage}
\end{table}

\clearpage
\begin{table}[h]
\caption[junk]{Journal of Echelle Observations}
\renewcommand{\footnoterule}{}
\begin{minipage}{5.75in}
\begin{tabular}{lccccc}
\hline \hline
WR       &   Slit      & Offset   &    Position      &  Exposure  & Date of \\
Star      &  Number   & from Star  &   Angle   &   (sec)    & Observation  \\
\hline
 Br 2     &   1    &  --            &  120$^\circ$  &  2$\times$900  &  96/1/9 \\
 Br 2     &   2    & 5$''$W         &  135$^\circ$  &  900           &  96/1/10 \\
 Br 2     &   3    & 4$''$N, 4$''$W &   45$^\circ$  &  900           &  96/1/11 \\
 Br 2     &   4    & 10$''$N, 35$''$W &    0$^\circ$  &  900           &  96/1/10 \\
 Br 13    &   1    &  --            &   90$^\circ$  &  900           &  96/1/10 \\
 Br 40a   &   1    &  --            &    0$^\circ$  &  900           &  96/1/11 \\
 Br 48    &   1    &  --            &   90$^\circ$  &  900           &  96/1/10 \\
 Br 52    &   1    &  --            &   90$^\circ$  &  600           &  96/1/10 \\
\hline
\end{tabular}
\end{minipage}
\end{table}

\clearpage

\begin{table}[h]
\caption[junk]{Ring Nebulae around Massive Stars}
\renewcommand{\footnoterule}{}
\begin{minipage}{5.75in}
\begin{tabular}{lllcllll}
\hline \hline
 Star      & Spectral & Nebula         &   Size    & V$_{\rm exp}$ &
 V$_{\rm exp}$/R & N/O & References.\\
Name     & Type     &   Name         &   (pc)    & (\kms) &
 (yr) & &\\
\hline
 Br 2   &  WN4     & in DEM\,L\,6   & 7$\times$4.5   & 16 & 1.8$\times10^5$&
0.070$\pm$0.015& 1,2\\
 Br 10  &  WC5     & in DEM\,L\,39  & 47$\times$25   & 42 & 4.3$\times10^5$& 
--& 3\\
 Br 13  &  WN8     & in DEM\,L\,56  & 10$\times$5.5  & 80 & 6.3$\times10^4$&
0.57$\pm$0.18& 1, 2\\
 Br 52  &  WN4+OB  &  DEM\,L\,240   & 25$\times$17.5 & 50 & 2.1$\times10^5$&
--& 1, 3\\
 Br 100 &  WN3-4   &  DEM\,L\,315   & 31$\times$22.5 & 47 & 2.8$\times10^5$&
--& 3\\
\hline
 WR 5   &  WN5     &  S\,308        & 14.4           & 60 & 1.2$\times10^5$&
1.66 & 4, 5\\
 WR 7   &  WN4     &  NGC\,2359     & 6.5            & 20 & 1.6$\times10^5$&
0.112& 5, 6\\
 WR 18  &  WN5     &  NGC\,3199     & 19$\times$15   & 20 & 4.3$\times10^5$&
0.148& 5, 7\\
 WR 40  &  WN8     &  RCW\,58       & 8$\times$6     & 110& 3.2$\times10^4$&
0.501& 5, 6\\
 WR 124 &  WN8     &  M\,1-67       & 2              & 42 & 2.4$\times10^4$&
2.95& 5, 8\\
 WR 136 &  WN6     &  NGC\,6888     & 4.2$\times$6.3 & 80 & 3.3$\times10^4$&
1.86& 5, 6\\
\hline
 AG Car &  LBV     &  AG Car nebula & 1.1$\times$1.0 & 70 & 7.1$\times10^3$&
5.74$\pm$2.25& 9, 10\\
 He3-519&  LBV     &  He3-519       & 2.2            & 62 & 1.8$\times10^4$&
-- & 11\\
HD\,168625& LBV    &  HD\,168625 nebula  & 0.06           & 20 & 3.0$\times10^3$&
--& 12 \\
HR Car   &  LBV    &  HR Car nebula & 1.3$\times$0.7 & $\sim100$ & 5.0$\times10^3$&
$>$3 & 13, 14\\
\hline
 R127   &  LBV     &  R127 nebula   & 1.9$\times$2.2 & 28 & 7.1$\times10^4$&
0.89$\pm$0.40& 15, 16 \\
 S119   &  LBV     &  S119 nebula   & 2.1$\times$1.9 & 25 & 8.0$\times10^4$&
1.41--2.45 & 16, 17\\
\hline
 Sk$-$69 279& O9f    &  Sk$-$69 279 nebula  & 4.5            & 14 & 1.6$\times10^5$&
-- & 18\\
 Sk$-$69 202& B3I    &  (SN 1987A's ring) & 0.4            & 10 & 2.0$\times10^4$&
1.6 & 19, 20\\
\hline
\end{tabular}

\null \null

REFERENCES. --
(1) This paper;
(2) Garnett \& Chu 1994;
(3) Chu 1983;
(4) Chu et al.\ 1982;
(5) Esteban et al.\ 1992;
(6) Chu 1988;
(7) Chu 1982;
(8) Solf \& Carsenty 1982;
(9) Smith 1991;
(10) Smith et al.\ 1997;
(11) Smith et al.\ 1994;
(12) Hutsemekers et al.\ 1994;
(13) Weis et al.\ 1997a;
(14) Nota et al.\ 1997;
(15) Appenzeller et al.\ 1987;
(16) Smith et al.\ 1998;
(17) Nota et al.\ 1994;
(18) Weis et al.\ 1997b;
(19) Crotts \& Heathcote 1991;
(20) Panagia et al. 1996.
\end{minipage}
\end{table}

\clearpage
\begin{center} {\large \bf Figure Captions} \end{center}

\figcaption{CTIO 0.9-m telescope CCD images of five small WR 
ring nebulae in the LMC.  The images are displayed with the
respective WR stars at the field center.  North is up and east 
is to the left.  Each panel covers a 2$'\times2'$ or 
4$'\times4'$ field of view, as indicated at its lower right 
corner.  The name of the WR star is given at the 
upper left corner and the filter information upper right 
corner of the panel.}

\figcaption{Long-slit echellograms of five small WR ring nebulae
in the LMC.  The name of the WR star, the slit offset (in units 
of arcsec) from the central WR star, and the position angle
(measured counterclockwisely from the north) of the slit are 
labeled above each panel.  The spatial orientation of the 
echellogram is marked at the upper right corner of the panel.
The wavelength increases to the right.  Each panel covers 
about 38.5 \AA, or 1760 km s$^{-1}$ at the H$\alpha$ line,
along the horizontal axis.  The nebular \ha\ and \nii\ lines
and the telluric ($\bigoplus$) \ha\ and OH lines are identified 
below the bottom panels.  The spatial extent of each panel, 
along the vertical axis, is 2$'$ for Br 2 and Br 13, and 4$'$ 
for Br 40a, Br 48, and Br 52.  The spatial scales are the same
as those for the images in Figure 1.}

\clearpage

\begin{figure}[tbh]
\centerline{\psfig{file=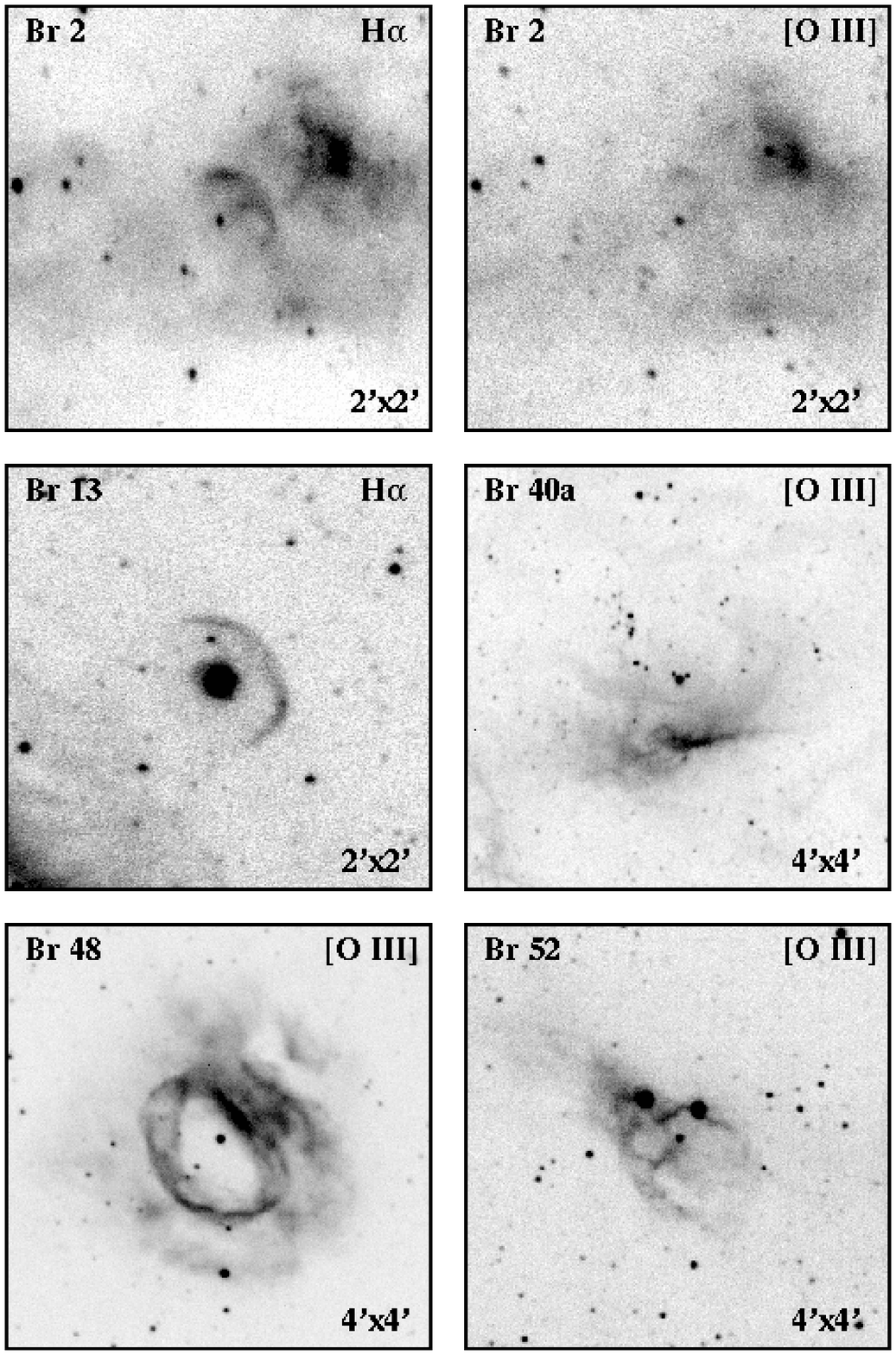}} 

\end{figure}
\clearpage

\begin{figure}[tbh]
\centerline{\psfig{file=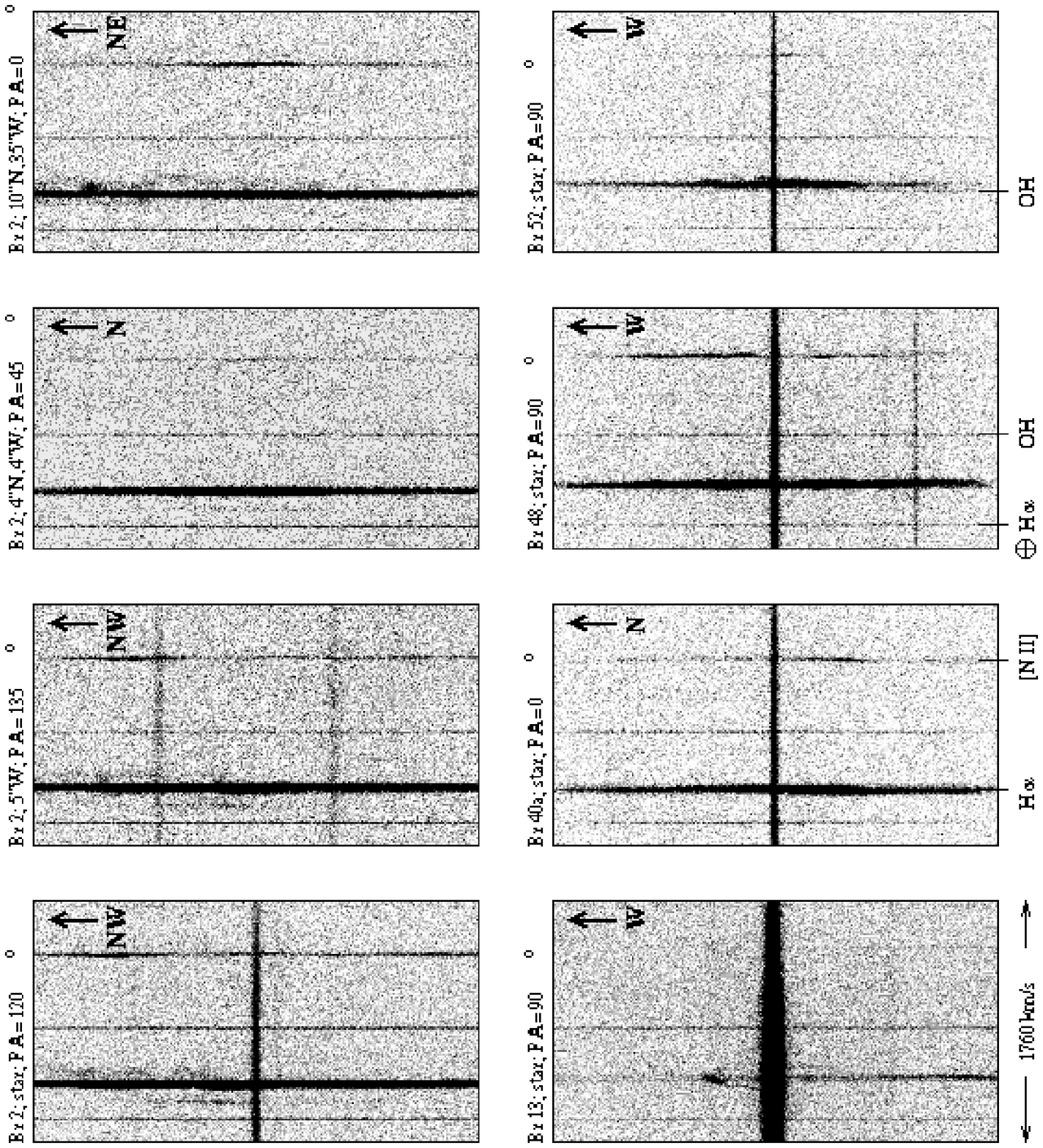}}
\end{figure}


\clearpage

\begin{references}
\reference{Ap87} Appenzeller, I., Wolf, B., \& Stahl, O. 1987, in 
Instabilities in Luminous Early-Type Stars, ed. H. J. G. L. M. Lamers \&
C. W. H. de Loore (Dordrecht: Reidel), 241
\reference{Br81} Breysacher, J. 1981, A\&AS, 43, 203
\reference{CMW75} Castor, J., McCray, R., \& Weaver, R. 1975, ApJ, 200, L107
\reference{Ch82} Chu, Y.-H. 1982, ApJ, 254, 578
\reference{Ch83} Chu, Y.-H. 1983, ApJ, 269, 202
\reference{Ch88} Chu, Y.-H. 1988, PASP, 100, 986
\reference{CG82} Chu, Y.-H., Gull, T. R., Treffers, R. R., Kwitter, K. B.,
\& Troland, T. H. 1982, ApJ, 254, 562
\reference{CK88} Chu, Y.-H., \& Kennicutt, R. C. Jr. 1988, AJ, 95, 1111
\reference{CL80} Chu, Y.-H., \& Lasker, B. M. 1980, PASP, 92, 730
\reference{CG83} Conti, P. S., \& Garmany, C. D. 1983, PASP, 95, 411
\reference{CH91} Crotts, A. P., \& Heathcote, S. R. 1991, Nature, 350, 683
\reference{DEM76}Davies, R. D., Elliott, K. H., \& Meaburn, J. 1976
MmRAS, 81, 89
\reference{Do94} Dopita, M. A., Bell, J. F., Chu, Y.-H., \& Lozinskaya, 
T. A. 1994, ApJS, 93, 455
\reference{Es92} Esteban, C., V\'{\i}lchez, J. M., Smith, L. J., \& Clegg, 
R. E. S. 1992, A\&A, 259, 629
\reference{GM95a} Garc\'{\i}a-Segura, G., \& Mac Low, M.-M. 1995a, 
ApJ, 455, 145
\reference{GM95b} Garc\'{\i}a-Segura, G., \& Mac Low, M.-M. 1995b, 
ApJ, 455, 160
\reference{GLM96} Garc\'{\i}a-Segura, G., Langer, N., \& Mac Low, 
M.-M. 1996, A\&A, 316, 133 (GLM96)
\reference{GML96} Garc\'{\i}a-Segura, G., Mac Low, M.-M., \& Langer,
N. 1996, A\&A, 305, 229 (GML96)
\reference{Ga98} Garnett, D. R. 1998, in IAU Symp. 190, New Views 
of the Magellanic Cloud, ed. Y.-H. Chu et al., in press
\reference{GC94} Garnett, D. R., \& Chu, Y.-H. 1994, PASP, 106, 626
\reference{Ha91} Haynes, R. F., et al. 1991, A\&, 252, 475
\reference{Hu94} Hutsemekers, D., Van Drom, E., Gosset, E., \& Melnick,
J. 1994, A\&A, 290, 906
\reference{JH65} Johnson, H. M., \& Hogg, D. E. 1965, ApJ, 142, 1033
\reference{La94} Langer, N., Hamann, W.-R., Lennon, M., Najarro, F.,
Pauldrach, A. W. A., \& Puls, J. 1994, A\&A, 290, 819
\reference{Ma94a} Marston, A. P., Chu, Y.-H., \& Garc\'{\i}a-Segura, G.
1994a, ApJS, 93, 229
\reference{Ma94b} Marston, A. P., Yocum, D. R., Garc\'{\i}a-Segura, G.,
\& Chu, Y.-H. 1994b, ApJS, 95, 151
\reference{MC93} Miller, G., \& Chu, Y.-H. 1993, ApJS, 85, 137
\reference{No94} Nota, A., Drissen, L., Clampin, M., Leitherer, C.,
Pasquali, A., Robert, C., Paresce, F., \& Robberto, M. 1994, in 
Circumstellar Media in the Late Stages of Stellar Evolution, ed. R. 
Clegg et al. (Cambridge: Cambridge Univ. Press), 89
\reference{No97} Nota, A., Smith., L. J., Pasquali, A., Clampin, M., 
\& Stroud, M. 1997, ApJ, 486, 338
\reference{Os96} Osterbrock, D. E., Fulbright, J. P., Martel, A. R., 
Keane, M. J., Trager, S. C., \& Basri, G. 1996, PASP, 108, 277
\reference{Pa91} Pakull, M. W. 1991, in IAU Symp. 143, Wolf-Rayet Stars
and Interrelations with Other Massive Stars in Galaxies, ed. K. A. van
der Hucht \& B. Hidayat (Dordrecht: Kluwer), 391
\reference{Pa96}Panagia, N., Scuderi, S., Gilmozzi, R., Challis, P. M., 
Garnavich, P. M., \& Kirshner, R. P. 1996, ApJ, 459, L17
\reference{Ro86} Rosado, M. 1986, A\&A, 160, 211
\reference{Sh83} Shaver, P. A., McGee, R. X., Newton, L. M., Danks, A. C.,
Pottasch, S. R. 1983, MNRAS, 204, 53
\reference{Sm91} Smith, L. J. 1991,  in IAU Symp. 143, Wolf-Rayet Stars
and Interrelations with Other Massive Stars in Galaxies, ed. K. A. van
der Hucht \& B. Hidayat (Dordrecht: Kluwer), 385
\reference{Sm94} Smith, L. J., Crowther, P. A., \& Prinja, R. K.
1994, A\&A, 281, 833
\reference{Sm98} Smith, L. J., Nota, A., Pasquali, A., Leitherer, C.,
Clampin, M., \& Crowther, P. A. 1998, ApJ, 503, 278
\reference{Sm97} Smith, L. J., Stroud, M. P., Esteban, C., V\'{\i}lchez,
J. M. 1997, MNRAS, 290, 265
\reference{SW83} Smith, L. J., \& Willis, A. J. 1983, A\&AS, 54, 229
\reference{SC82} Solf, J., \& Carsenty, U. 1982, A\&A, 116, 54
\reference{SSW75} Steigman, G., Strittmatter, P. A., \& Williams, R. E.
1975, ApJ, 198, 575
\reference{Hu81} van der Hucht, K. A., Conti, P. S., Lundstrom, I., \&
Stenholm, B. 1981, Space Sci. Rev., 28, 227
\reference{We77} Weaver, R., McCray, R., Castor, J., Shapiro, P., \&
Moore, R. 1977, ApJ, 218, 377
\reference{We97a} Weis, K., Duschl, W. J., Bomans, D. J., Chu, Y.-H., \&
Joner, M. D. 1997a, A\&A, 320, 568
\reference{We97b} Weis, K., Chu, Y.-H., Duschl, W. J., Bomans, D. J. 1997b,
A\&A, 325, 1157

\end{references}
\end{document}